\documentclass[12pt]{article}

\usepackage{color}

\begin{document}

\title{{\bf Scale Invariant Hairy Black Holes}}

 \author{M\'aximo Ba\~nados$^{(a)}$ and Stefan Theisen$^{(b)}$ \\ \\
{\it $^{(a)}$Departamento de F\'{\i}sica, P. Universidad Cat\'olica de Chile},\\
{\it Casilla 306, Santiago 22,Chile.} \\
{\it $^{(b)}$Max-Planck-Institut f\"ur Gravitationsphysik,}\\
{\it Albert-Einstein-Institut, 14476 Golm, Germany}\\
{\tt mbanados@fis.puc.cl, \  theisen@aei.mpg.de }}
\maketitle

\def\Snote#1{\footnote{\bf S:~#1}}
\def\Mnote#1{\footnote{\bf M:~#1}}

\begin{abstract}
Scalar fields coupled to three-dimensional gravity are considered.  We uncover a scaling
symmetry present in the black hole reduced action, and use it to prove a Smarr formula
valid for any potential. We also prove that non-rotating hairy black holes exists only
for positive total energy.  The extension to higher dimensions is also considered.
\end{abstract}

\section{Introduction and discussion}

The system of Gravity coupled to scalar fields has recently been under considerable
scrutiny. Asymptotically AdS Hairy black holes have been shown to exist in
\cite{HMTZ1,HMTZ2,HeMa1,MTZ}. The issue of defining meaningful conserved charges has
been considered in \cite{HMTZ1,HMTZ2,Bar,HeMa1,HolIshMar,PaSk}. Earlier references
include \cite{AbbDes,AshMag,HenTei,Tor}. Within the AdS/CFT correspondence, the coupling
of scalar matter was considered in \cite{dHSS,BFS}.

Our aim in this paper is to make some general remarks on the structure of hairy black
holes in three dimensions. Our key ingredient is the existence of a scale symmetry in
the reduced action governing the black hole ansatz. This symmetry exists for any
potential $V(\phi)$ and provides, via Noether's theorem, a radially conserved charge. We
use this charge to find a relationship between the black hole parameters at infinity
with those at the horizon.

Our main result is the following. Let $M$, $J$ and $S$ be the total energy, angular
momentum, and entropy of a black hole solution with some non-zero scalar field $\phi$.
Let $T$ and $\Omega$ be the black hole's temperature and angular velocity. Assuming that
the matter field is finite at the horizon and vanishes at infinity, it follows that
these parameters must satisfy the three-dimensional Smarr \cite{Smarr} relation,
\begin{equation}\label{SmarrJ}
M = {1 \over 2} TS - {1\over2}\Phi Q-\Omega J .
\end{equation}

The remarkable aspect of this result is its universality. In fact the scalar field and
its potential play no role. The only condition on the matter field is that it must be
finite everywhere, and zero asymptotically. Of course this imposes non trivial
constraints on the class of potentials being considered, which must elude the no-hair
theorems. But, if the black hole exists, then it must satisfy (\ref{SmarrJ}).

The first law of black hole thermodynamics,
\begin{equation}\label{flJ}
\delta M = T \delta S -\Phi\delta Q - \Omega\delta J,
\end{equation}
is also valid in this theory. Inverting the Smarr relation (\ref{SmarrJ}) one finds that
$S(M,J,Q)$ must be a homogeneous\footnote{I terms of the total mass, the homogeneity
property reads $M(\sigma S,\sigma^2 J,\sigma^2 Q) = \sigma^2 M(S,J,Q) $} function of degree 1/2 of
its arguments, $S(\sigma^2 M,\sigma^2 J,\sigma Q)=\sigma S(M,J,Q)$. This is certainly true for
the vacuum BTZ back hole. Our result implies that hairy black holes, regardless of the
potential chosen, satisfy the same scaling relation.

A remark is in order here: the homogeneity property of $S(M,J,Q)$ is not a consequence of
simple dimensional analysis and scaling arguments as is the case e.g. for the
Kerr-Newman metrics, c.f. \cite{GP}. This is due to the presence of an additional
dimensionful parameter, the curvature radius of the AdS space-time or, equivalently, the
cosmological constant\footnote{Generalized Smarr relations have been considered in
\cite{Bar2}. }. The reason why (\ref{SmarrJ}) holds nevertheless, even in the presence
of scalar hair, is the scaling symmetry and the associated radially conserved charge.

In the non-rotating neutral case, $J=Q=0$, we can use (\ref{SmarrJ}) and (\ref{flJ}) to find the
general expression for the temperature of non-rotating black holes,
\begin{equation}\label{TM}
T = \kappa\, M^{1/2}
\end{equation}
where $\kappa$ is a constant with no variation.  This means, in particular, that for any
potential $V(\phi)$ the specific heat of the black hole is positive.

It is interesting to compare (\ref{TM})  with the result reported in \cite{HMTZ1}. In
three dimensions, \cite{HMTZ1} considered the potential,
\begin{equation}\label{Vnu}
V = -{1 \over 8} \left( \cosh^6\phi + \nu \sinh^6\phi \right),
\end{equation}
where $\nu$ is a real parameter. An explicit black hole configuration was displayed,
whose temperature as a function of the total energy follows the general law (\ref{TM}),
and $\kappa$ becomes a complicated function of the parameter $\nu$.

The three dimensional structure can be generalized to higher-dimensional black holes
with toroidal topology \cite{Ami,Vanzo}, as well as to black holes on flat branes
\cite{HorRoss,Witten}. This is analyzed in Sec. \ref{TorusSec}.

After this work was completed we became aware of \cite{Clement} in which the Smarr
relation for hairy black holes in three-dimensions was also found. The parametrization
of the reduced action used in this reference is very different from ours, and the
relevant symmetry is an $SL(2,R)$ group rather than the scaling symmetry which we have
employed.

%--------------------------------------
\section{Reduced action in d=3 and scaling symmetry}
\label{RedAction}

Consider the action describing three-dimensional gravity coupled to a scalar field
$\phi$,
\begin{equation}
I ={1 \over 16\pi G} \int \left( R - 8 g^{\mu\nu} \partial_\mu\phi \partial_\nu \phi -
16  V(\phi) \right)\sqrt{-g}\,d^3 x. \label{action}
\end{equation}
We assume that $V(\phi)$ has a non-zero negative value at $\phi=0$, such that the
gravitational background is anti-de Sitter space.

We shall first consider non-rotating solutions. The generalization can be done
straightforwardly and will be indicated in Sec. \ref{AngularMomentum}.  Consider static,
spherically symmetric solutions of the form
\begin{equation}\label{metric}
ds^2 =  -\gamma(r)^2 h(r) dt^2 + {dr^2 \over h(r)} + r^2 d\varphi^2,\ \ \ \ \  \phi =
\phi(r)\,.
\end{equation}
Solutions of this form include, for example, black holes and soliton solutions. The
solitons are relevant for AdS/CFT applications, as recently considered in \cite{HeHo}.
We shall concentrate in this paper on black holes. One can write a reduced action for
this problem,
\begin{equation}\label{Ired}
I[h,\gamma,\phi]=-{(t_2-t_1)\over 8G}\int dr\,\gamma\left( h'+8r
h\phi'^2+16rV(\phi)\right)+B .
\end{equation}
where $B$ is a boundary term that we shall consider below. The equations of motion are,
\begin{eqnarray}
  h' + 8rh \phi'^2  + 16r V &=& 0\,, \nonumber\\
  -\gamma' + 8r\gamma \phi'^2 &=& 0\,,  \label{3deq}\\
  -(r\gamma h\phi')' + r\gamma V_{,\phi}&=&0\,. \nonumber
\end{eqnarray}
They can be shown to be consistent with the original Einstein equations.

The key observation is that the action (\ref{Ired}) is invariant under the scale
transformations\footnote{In the matter-free case this scale symmetry was already
observed in \cite{HoHu}.}
\begin{eqnarray}
  \tilde r &=& \sigma r \label{dr}\,,\\
  \tilde h(\tilde r) &=& \sigma^2 \, h(r)\,,  \label{dh1}\\
  \tilde \gamma(\tilde r) &=& \sigma^{-2} \gamma(r)\,, \label{dg1} \\
  \tilde \phi(\tilde r) &=& \phi(r)  \label{dphi0} \,,\ \ \ \
\end{eqnarray}
with $\sigma$ a (positive) constant.

By direct application of Noether's theorem to the above symmetry one finds that the
combination
\begin{eqnarray}
C = \gamma\left( - h  + {1 \over 2} r  h' + 8r^2 h\, \phi'^2 \right) \label{C}
\end{eqnarray}
is conserved, $C'=0$. One can in fact prove this directly from the equations
(\ref{3deq}). A crucial property of this conservation law is that it holds for any
potential $V(\phi)$. This will allow us to make general statements about the nature of
3d black holes coupled to scalar fields.

Our strategy is now the following. Since $C$ does not depend on $r$ we can use it to
find a relationship between the asymptotic parameters $M,\beta$ and the horizon $r_+$.
As we shall see, this relation is precisely the Smarr relation (\ref{SmarrJ}). But
before we can state this result, we need to find an expression for the energy of this
system.

%=========================================
\section{Energy and entropy}

The analysis in this section assumes a generic potential. For some specific cases, as
masses saturating the BF bound, a separate analysis may be needed.

The boundary term $B$ that appears in (\ref{Ired}) is fixed by the condition that, upon
varying the action, all boundary terms cancel for a set of given boundary conditions. At
this point we shall switch to the Euclidean formalism, and interpret the on-shell action
as the free energy of the thermodynamical system \cite{GH}. The Euclidean action $I_{\rm
E}$ is the same as (\ref{Ired}), except that $(t_2-t_1)=1$ and an overall sign, such
that the weight in the functional integral is $e^{-I_{\rm E}}$.

In the Euclidean Hamiltonian formalism, the boundary consist of two disconnected pieces,
one in the asymptotic region $r\rightarrow\infty$ and the other at the horizon.  The
boundary term $B$ is specified by the condition,
\begin{eqnarray}\label{deltaB}
\delta B &=& - {1 \over 8G} \, \gamma\, ( \delta h + 16 r h \phi' \delta \phi
)|_{r=\infty} + {1 \over 8G} \, \gamma\, \delta h\, |_{r=r_+}\,,
\end{eqnarray}
where the horizon is defined by the equation $h(r_+)=0$. We assume that all fields are
regular there.

Assuming that the matter field vanishes at infinity, Eqns. (\ref{3deq}) imply that,
asymptotically, $\gamma'=0$. We write $\gamma(\infty)=\beta$, where $\beta$ is a
constant equal to the Euclidean period at infinity\footnote{Note that solutions of the
form $\gamma \sim \log(r)$ will not occur for a generic potential.}.

The boundary term now has the form $\delta B = \beta \delta M - \delta S$ where the
variation of mass and entropy are given by,
\begin{eqnarray}
\delta M &=&  -{1 \over 8G}( \delta h + 16 r h \phi' \delta \phi )|_{r=\infty}\,, \label{dM}\\
\delta S &=&  -{1 \over 8G} \, \gamma\, \delta h\,  |_{r=r_+}\,,  \label{dS}
\end{eqnarray}
As usual in the Hamiltonian formalism the entropy comes from the variation of the action
at the horizon\cite{BrMar,BTZ,Wald}. Our task now is to identify the actual values of
$S$ and $M$.

The boundary term at the horizon gives the usual Bekenstein-Hawking entropy without any
modifications.  In fact, from $h(r_+)=0$ and $(h+\delta h)(r_+ +\delta r_+)=0$ it
follows that $\delta h(r_+)=-h'(r_+)\delta r_+$, as long as $h'(r_+)\neq 0$, which is
satisfied for non-extreme black holes. In addition, the value of $\gamma$ at the horizon
cannot be arbitrary. To avoid conical singularities at $r=r_+$ one must impose\cite{GH}
 \begin{equation}\label{cs}
 \gamma(r_+) h'(r_+) = 4\pi\,.
 \end{equation}
These two conditions allow us to identify $S$ as
\begin{equation}\label{S}
S = {2\pi r_+ \over 4G},
\end{equation}
just as in the matter-free system.

We now turn to the problem of integrating (\ref{dM}) to extract the value of $M$. This
problem is more subtle because we have not specified the potential. We shall integrate
(\ref{dM}) by using again the scale invariance discussed above, which maps solutions to
solutions.

The idea is the following. The functions $h$ and $\phi$ have scaling dimensions 2 and 0,
respectively. From (\ref{dM}) we conclude that $M$ must have scaling dimension 2. This
means that under the scale variations of $h$ and $\phi$,
\begin{eqnarray}
  \delta h &=& \delta\sigma (-rh' + 2h), \label{dh11} \\
  \delta \phi &=& -r\delta\sigma\phi', \label{dphi11}
\end{eqnarray}
the corresponding variation of $M$ satisfies
 \begin{equation}\label{M2}
\delta M = 2\delta\sigma\, M.
\end{equation}
We now replace (\ref{dh11}) and (\ref{dphi11}) in (\ref{dM}) and, comparing with
(\ref{M2}), we obtain the desired formula for $M\,$\footnote{The relationship between
asymptotic functional variations and scale transformations can be checked explicitly in
some examples. For the BTZ black hole with $h(r) = r^2 + h_0$ one has $\delta h = \delta
h_0$. The constant $h_0$ has scaling dimension 2, $\delta h_0 = 2\delta\sigma h_0 $. One
can check that in fact $\delta\sigma (-rh' +2h)=\delta h_0$. In the system studied in
\cite{HMTZ1}, the asymptotic solution is $h \simeq r^2+4B r-3(1+\nu)B^2$. It is direct
to check that $\delta h = \delta \sigma(-rh'+2h)$ with $\delta B = \delta\sigma\, B$, as
claimed. Note finally that this correspondence fails in higher dimensional gravity. See
Sec. \ref{TorusSec} for details on this case.},
\begin{equation}\label{M}
M = {1 \over 8G} \left( - h  + {1 \over 2} r  h' + 8r^2 h\, \phi'^2 \right).
\end{equation}
Before explaining and discussing the validity of this formula let us check that it gives
the right results in known cases. For a BTZ black hole, $h = r^2 - 8Gm$ and $\phi=0$.
One finds $M=m$, as expected. A less trivial example is the exact hairy black hole
solution found in \cite{HMTZ1} with $h = r^2 + 4Br -3(1+\nu) B^2 + {\cal O}(1/r)$ and
$\phi = (B/r)^{1/2} - 2/3 (B/r)^{3/2} + {\cal O}(1/r^{5/2})$. Replacing this field in
(\ref{M}) one obtains
\begin{equation}
M = {3(1+\nu)B^2 \over 8G}\,,
\end{equation}
in full agreement with \cite{HMTZ1}.

Now, some comments on the derivation and validity of (\ref{M}) are necessary. The
variations (\ref{dh11},\ref{dphi11}) do not explore the full set of asymptotic
solutions. In fact, (\ref{dh11}) and (\ref{dphi11}) represent a 1-parameter ($\sigma$)
set of variations. On the other hand, the equations are of first order for $h(r)$, and
second order for $\phi(r)$ and the full space of solutions has three parameters. The key
step is that since $M$ is a ``function of state" (exact differential), its value does
not depend on the path chosen and in this sense the formula (\ref{M}) is the correct
one. However, we must now make sure that $\delta M$, as given in (\ref{dM}), is actually
an exact differential. An equivalent way of stating this is that the existence of a
well-defined variational principle requires $B$ in (\ref{Ired}) to exist, not just
$\delta B$.

We do not need to worry about the first term in (\ref{dM}), $\delta h$, which is exact.
The second piece, $rh\phi'\delta\phi$, needs a separate analysis. For a generic
potential the  asymptotic form of the scalar field on AdS is,
\begin{equation}\label{phiinf}
\phi =   {a \over r^{\triangle_-}}  + {b \over r^{\triangle_+}} + \cdots\,,
\end{equation}
where, for static black holes, $a$ and $b$ are arbitrary constants and represent the two
degrees of freedom associated to $\phi$. The exponents $\triangle_\pm$ are the solutions
to a quadratic equation and satisfy $\triangle_+ + \triangle_-=2.$  We assume that both
are positive.

Plugging (\ref{phiinf}) into (\ref{dM}) one finds finite terms of the form $f(a,b)\delta
a+g(a,b)\delta b$. In order to write these terms as total variations (to achieve path
independence) one needs to assume a relationship between $a$ and $b$. This restriction
on the space of solutions is generic and was also found in
\cite{HMTZ1,HMTZ2,HeMa1,HeMa2}.

The particular choice considered in these references (generalized here to arbitrary
$\triangle_\pm)$ is
\begin{equation}\label{k}
b = \eta\, a^{{\triangle_+ \over \triangle_-}} , \ \ \ \ \ \ \ \     \delta \eta =0\,,
\end{equation}
where $\eta$ is held fixed. This choice is consistent with the full anti-de Sitter
asymptotic group, although this will not be relevant for our discussion\footnote{Note
that this particular choice is by no means the most general. For solitonic solutions, as
in \cite{HeHo}, $a$ and $b$ become related in a different way. We shall consider
solitons in this theory elsewhere.}.

For our purposes, the choice (\ref{k}) is singled out by demanding scale invariance of
the asymptotic solution.  In fact, once a relationship between $a$ and $b$ is assumed,
the only function $b=b(a)$ consistent with (\ref{dphi11}) is precisely (\ref{k}). We
conclude that on the space of solutions satisfying the boundary conditions (\ref{k}),
the formula (\ref{M}) for $M$ is correct.

Finally, we point out that the remarkable cancelations of divergent pieces in the total
mass $M$, discovered in \cite{HMTZ1,HMTZ2,HeMa1}, can be seen in this case from a
different perspective. Note that, up to the factor $\gamma(r)$ which becomes a constant
at infinity, $M$ is exactly equal to the scale charge $C$ displayed in (\ref{C}).  Since
$C$ does not depend on $r$, it cannot diverge; the total mass is then finite.

%---------------------------
\section{ Thermodynamics of the hairy black hole}

%-------------------------------------------------
\subsection{The First Law }

The first law for our class of black hole solutions can be checked by standard
Hamiltonian arguments. The form of the action, derived in the previous section, after
all boundary terms have been included is,
\begin{equation}\label{Ifl}
I[\beta] = \int dr\, \gamma\, {\cal H} \ + \ \beta\, M \ - \ S(r_+),
\end{equation}
where $M$ is given in (\ref{M}) and $S$ is the usual entropy in three dimensions, given
in (\ref{S}). ${\cal H}=0$ is one of the equations of motion. By construction, this
action has an extremum when evaluated on solutions with $\beta$ fixed. The on-shell
value of $I$ only depends on $\beta$. The value of $M$ is such that $I$ has an extremum.

Since the bulk contribution is proportional to a constraint, the on-shell value of the
action is
\begin{equation}\label{Ios}
I[\beta] = \beta \, M - S(M)
\end{equation}
where $r_+$ is written as a function of the total energy $M$ using the
solution\footnote{There is a non-trivial assumption here, namely, that $r_+$ depends
only on $M$, and not on the scalar field parameter $a$. We prove in the next section
that black holes exists only for special values of $a=a(M)$, and hence $a$ is not an
independent parameter.}. $M$ is not fixed but has to be chosen such that $I$ has an
extremum, that is,  the first law is satisfied,
\begin{equation}
\beta \delta M = \delta S.
\end{equation}

%------------------------------
\subsection{The Smarr relation}

We are now ready to prove our main result. We go back to the expression for the scaling
charge $C$ given in (\ref{C}). Comparing (\ref{C}) with (\ref{M}) we conclude that the
scaling charge is proportional to the total mass. Evaluating $C$ at infinity we get the
exact relation,
\begin{equation}\label{CM}
C = 8 \beta MG.
\end{equation}
On the other hand, since $C$ is $r-$independent we can also evaluate it at the horizon
$h(r_+)=0$ to get
\begin{equation}
C = 2\pi r_+ \,.
\end{equation}
Here we have used the condition of absence of conical singularities (\ref{cs}).
Comparing the values of $C$ at infinity and at the horizon we find the equation,
\begin{equation}\label{rel}
\beta M  = {1 \over 2}\, {2\pi r_+ \over 4G}
\end{equation}
representing the non-rotating version of (\ref{SmarrJ}). This relation is satisfied for
any black hole solution with or without scalar field. Of course, this is also true for
the BTZ vacuum black hole, as can be readily checked. The rotating case, leading to
(\ref{SmarrJ}), will be indicated in Sec. \ref{AngularMomentum}.

%-------------------------------
\subsection{Positivity of energy}

We prove now that a hairy black hole can exist only if the total mass $M$ is positive.
To see this we first note that the field $\gamma(r)$ does not change sign in the whole
range $r_+\leq r \leq \infty$. In fact, directly from the equations of motion
(\ref{3deq}) we can write the formal solution
\begin{equation}
\gamma(r) = \gamma_0\, e^{\int^r ds\, 8s \phi'(s)^2 }
\end{equation}
where $\gamma_0$ is an arbitrary integration constant. This expression for $\gamma$ is
manifestly positive, if $\gamma_0$ is positive. Now, the scaling charge evaluated at the
horizon and at infinity gives the equation (we relax here the condition (\ref{cs}) and
consider either Minkowskian or Euclidean signature)
\begin{equation}
16 G \gamma_\infty\, M = \gamma_+  h'_+ \, r_+
\end{equation}
where the subscript $_+$ indicates the corresponding function evaluated at $r_+.$ The
function $h(r)$ must be positive outside the horizon, and vanishes at $r_+$. This means
that $h'_+ >0$. Since $\gamma(r)$ does not change sign and $r_+$ is positive, we
conclude that this equation requires
\begin{equation}
M>0.
\end{equation}

%----------------------------
\subsection{Temperature and specific heat}

Combining (\ref{rel}) with (\ref{S}) and the first law we derive the general relation
\begin{equation}\label{kappa}
8MG = \kappa_0^2\, r_+^2\,,
\end{equation}
where $\kappa_0$ is an arbitrary (dimensionless) integration constant, with no
variation. This constant cannot be computed from this analysis and depends on the
details of the potential, as well as all other fixed parameters. For example, for the
BTZ black hole $\kappa_0=1$ while for the potential (\ref{Vnu}) considered in
\cite{HMTZ1} one finds that $\kappa_0$ becomes a complicated function of $\nu$.

All thermodynamical properties can now be extracted, for example the temperature as a
function of the mass gives,
\begin{equation}
{1 \over \beta} = T = \kappa_0 { \sqrt{2 MG} \over \pi}\,,
\end{equation}
as announced. We can also check that the specific heat,
\begin{equation}
c = {\partial M \over \partial T} = {2\pi r_+ \over 4G},
\end{equation}
becomes equal to the entropy (this was also noted by \cite{HMTZ1} in their particular
example).

%=================================
\section{A closer look at the hairy black hole}

A hairy black hole is, by definition, a solution to the Einstein + matter system
displaying a regular horizon. In particular, the value of the matter field $\phi(r)$ at
the horizon  must be finite.  We have argued in the previous section that, if the black
hole exist, then the Smarr (\ref{SmarrJ}) relation is satisfied. However, we have said
very little about the conditions for the existence of a black hole.

The condition of regularity at the horizon imposes constraints on the solutions which
can be analyzed using scale invariance. In this section we will prove that for a given
value of $\eta$ (see (\ref{k})), the values of the total mass $M$ and the parameter $a$
have to be fine tuned in order to have a regular black hole.  This means that, apart
from $\eta$ which acts as an external parameter with no variation, the only free
parameter in the black hole spectrum is the total energy $M$.

Consider the set of equations of motion (\ref{3deq}). We would like to find a solution
displaying a regular event horizon $h(r_+)=0$.  At the point $r=r_+$, the matter field,
and its derivatives up to some sufficiently high order, must be finite. In particular,
\begin{equation}
\phi(r_+) = \phi_0.
\end{equation}
Define the new field
\begin{equation}\label{chi}
\chi(r) = \phi(r) - \phi_0
\end{equation}
which vanishes at the horizon, $\chi(r_+)=0$. Near the horizon the new field $\chi$ is
small and hence, without specifying $V(\phi)$, we write the near horizon series,
\begin{equation}\label{Vseries}
V(\chi) = v_0 + v_1\, \chi + v_2\, \chi^2 + \cdots
\end{equation}
where the constants $v_i$ depend on the potential $V$ and $\phi_0$.

Under these conditions, the fields $h,\gamma,\phi$ have the following series expansions
near the horizon
\begin{eqnarray}
h      &=& h_1 (r-r_+) + h_2 (r-r_+)^2 + \cdots \\
\gamma &=& \gamma_0 + \gamma_1 (r-r_+) + \gamma_2 (r-r_+)^2 + \cdots   \label{Hseries}\\
\chi   &=& \chi_1 (r-r_+) + \chi_2 (r-r_+)^2 + \cdots
\end{eqnarray}
Recall that in the Euclidean formalism the values of $h'$ and $\gamma$ at $r=r_+$ are
linked by (\ref{cs}), that is $h_1\, \gamma_0=4\pi$.  Our conclusions, however, do not
depend on the signature.

We have assumed that no fractional powers or logs are present because they would induce
divergences in the derivatives of the fields.

We now plug this series expansion into the equations of motion and solve for the
coefficients order by order. This is a straightforward exercise that we do not display
here. The important comment is that all coefficients are fixed in terms of $\phi_0$ and
$r_+$ (recall that $\phi_0$ enters in the coefficient $v_0$ in the series
(\ref{Vseries}) for the potential). There are thus 2 arbitrary constants at the horizon:
\begin{equation}
\mbox{Horizon data:   }\{\phi_0\, , \, r_+\},
\end{equation}
as opposed to the series analysis at infinity with,
\begin{equation}
\mbox{Asymptotic data:   }\{\eta, \  a\, , \, M\}.
\end{equation}
What happens here is that the series expansion (\ref{Hseries}) is not the most general
one.  There exists other solutions with logs or fractional powers (probably depending on
the potential), which are not contained in the regular ansatz. \footnote{This has also
been remarked in \cite{Gubser}.}

We conclude that if one integrates from infinity to the horizon, the values of $a,\eta$
and $M$ must be fine tuned in order to reach a regular event horizon. Conversely, if one
integrates from the horizon, prescribing the values of $r_+$ and $\phi_0$, one gets at
infinity a surface in the $\eta,a,M$ space. Actually, we can say something else. We
shall prove now that $\eta$ only depends on the value of $\phi_0$, and not on $r_+$,
\begin{eqnarray}
  \eta = \eta(\phi_0).
\end{eqnarray}
To see this, suppose we are given a solution to the equations of motion,
$h(r),\gamma(r),\phi(r)$ displaying a regular event horizon.  Using scale invariance we
can provide immediately another exact solution to the equations by the simple
transformation
\begin{eqnarray}
  \tilde h(r) &=& \sigma^2\, h(r/\sigma) \\
  \tilde \gamma(r) &=& \sigma^{-2}\, \gamma(r/\sigma) \\
  \tilde \phi(r) &=& \phi(r/\sigma)
\end{eqnarray}
The new solution is a different one! If the horizon in the first solution was at
$r=r_+$, then the location of the horizon in the second solution is at
\begin{equation}
\tilde r_{ +} = \sigma r_+.
\end{equation}
In fact, $\tilde h(\tilde r_+)=0.$ This means that acting with scale transformations, we
can cover all possible values of $r_+$. On the other hand, the value of $\phi_0$ remains
unchanged since
\begin{equation}
\tilde \phi(\tilde r_+) = \phi (r_+) = \phi_0.
\end{equation}

Acting with scale transformations we thus cover all solutions with a given value of
$\phi_0$. Now, scale transformations act on the asymptotic parameters leaving $\eta$
invariant. We thus conclude that the asymptotic parameter $\eta$ is in one-to-one
correspondence with the value of $\phi$ at the horizon
\begin{equation}
\phi_0 \ \ \ \Leftrightarrow \ \ \  \eta.
\end{equation}
For a given value of $\eta$, the value of $\phi_0$ is determined. In the example of
\cite{HMTZ1}, $\eta=-2/3$ and $\phi_0=\tanh^{-1}(1/\sqrt{3}).$

Recall that $\eta$ is fixed in the action principle, and acts as an external parameter.
For fixed $\eta$ (and hence $\phi_0$), the remaining degrees of freedom are $M$ and $a$,
at infinity, and $r_+$ at the horizon. This means that if one integrates from the
horizon, varying the values of $r_+$, one obtains at infinity a curve in the $M,a$
plane. As we have shown this curve will cover only the $M>0$ half plane. Of course, for
different values of $\eta$, the curve changes.

%===============================================
%===================================================
\section{Adding angular momentum}
\label{AngularMomentum}

We will now extend the discussion of the thermodynamics to black holes with angular
momentum. This requires a change of the ansatz for the metric (\ref{metric}) to
\begin{equation}\label{metricJ}
ds_{\rm E}^2 =\gamma(r)^2 h(r) dt^2+{dr^2\over h(r)}+r^2(d\varphi+n(r)dt)^2,\ \ \ \ \
\phi=\phi(r)\,.
\end{equation}
The reduced action is
\begin{equation}\label{IredJ}
I[h,\gamma,n,\phi]={1\over 8G}\int dr\left\lbrace\gamma\Big(-{2p^2\over r^3} +h'+8r
h\phi'^2+16 r V\Big)+2 n p'\right\rbrace+B\,.
\end{equation}
Here $p=\pi^r_\varphi=-{r^3\over2\gamma}n'$. The bulk term of the action vanishes
on-shell. The equation of motion for $n$ gives $p={\rm const.}$. By shifting the angular
coordinate we arrange for $n(r_+)=0$.

The action is invariant under (\ref{dr}-\ref{dphi0}), augmented by
\begin{eqnarray}
\tilde p(\tilde r) & = & \sigma^2 p(r)  \label{dp}\,, \\
\tilde n(\tilde r) & = & \sigma^{-2}n(r)\label{dn}\,.
\end{eqnarray}
This leads to the following radially conserved Noether charge
\begin{equation}\label{CJ}
C=\gamma\left(- h+{1\over2} h'r+8 r^2 h\phi'^2\right)-2 n p\,.
\end{equation}
One checks that indeed $C'=0$ by virtue of the equations of motion.

The boundary terms $B$ must again be chosen such that $\delta B$ cancels the boundary
terms which appear when one extremizes the action. One finds
\begin{eqnarray}\label{deltaBJ}
\delta B & = & -{1\over 8 G}\Bigl\lbrace(\beta(\delta h+16 r h \phi'\delta\phi)
+2n\delta p\Bigr\rbrace\Big|_{r=\infty}
+{1\over 8G}(\gamma\delta h +2n\delta p)\Bigl|_{r=r_+}\nonumber\\
& \equiv & \beta(\delta M+\Omega \delta J)-\delta S
\end{eqnarray}
Here we have used that $h(r_+)=0$ and the definitions $\gamma(\infty)\equiv\beta,
\,n(\infty)\equiv\beta\Omega$. It follows from the equations of motion that $\beta$ and
$\Omega$ are finite. The first two terms are the contribution from $r=\infty$, the last
term is the contribution from the horizon. Replacing once more the functional variations
by those which follow from the scaling properties of the fields, combined with the fact
that $M$ and $J$ have weight two, one finds
\begin{equation}\label{CMJ}
C=8G\beta (M+\Omega J)
\end{equation}
From the contribution at $r=r_+$ we find again Eq.(\ref{S}), i.e. $S={2\pi r_+\over
4G}$.

In order to find a relation between $M, J$ and $S$, we use the fact that $C$ is a
constant. While its expression at $r=\infty$ was used to relate it to $M$ and $J$, we
now use its expression at the horizon to relate it to $S$. This leads to
\begin{equation}\label{relJ}
\beta (M+\Omega J)={1\over2}S\,.
\end{equation}
as promised.  One easily verifies that this relation is satisfied for the rotating BTZ
black hole.

%===============================================
%===================================================
\section{Adding electric charge}
\label{ElectricCharge}

In this section we will generalize the previous discussion to rotating electrically
charged black holes in the presence of a charged scalar field.

Starting point is the covariant action
\begin{equation}\label{IcovQ}
I={1\over16\pi G}\int d^3x\sqrt{-g}(R-16 g^{\mu\nu}D_\mu\phi(D_\nu\phi)^*-16V-4 F^{\mu\nu}F_{\mu\nu})
\end{equation}
with $D_\mu\phi=\partial_\mu+iq A_\mu\phi$ where $q$ is the electric charge of thes calar field.
For the metric and the scalar field we make again the ansatz eq.(\ref{metricJ}), for the gauge field
we also assume $A_\mu=A_\mu(r)$. This leads to the reduced euclidean action
\begin{eqnarray}
I^{\rm red}_{\rm E}&=&{1\over 8G}\int dr\biggl\lbrace{r^3 n'^2\over 2\gamma}+h'\gamma+16 r\gamma V
+{16\over r}q^2\gamma|\phi|^2 A^2+16r h\gamma|\phi'+iqA_r\phi|^2\nonumber\\
&&\qquad +{8\over r}h\gamma A'^2
+{16r\over\gamma h}q^2|\phi|^2(nA-\Phi)^2+{8r\over\gamma}(\Phi'-nA')^2\biggr\rbrace+B
\end{eqnarray}
Here we have introduced the notation $\Phi=A_t,\,A=A_\varphi$.
We have checked that the equations of motion derived from this reduced action imply
those derive from the covariant action.

It is convenient to introduce the following momenta:
\begin{eqnarray}\label{defpE}
p &=& -{1\over 2\gamma}r^3 n'\\
{\cal E}&=&{16r\over\gamma}(\Phi'-n A')\nonumber\\
\end{eqnarray}
where, as before, $p=\pi^r_\varphi$ whereas ${\cal E}$ is proportional to the radial
component of the electric field.
In terms of these the reduced action reads
\begin{eqnarray}\label{IredQ}
I^{\rm red}_{\rm E}&=& \int dr\biggl\lbrace\gamma\Bigl[-{2p^2\over r^3}+h'+16r V+16rh|\phi'+iq A_r\phi|^2
+{16\over r}q^2 A^2|\phi|^2+{8h\over r}A'^2-{1\over32 r}{\cal E}^2\Bigr]\nonumber\\
&&\qquad+{16r\over\gamma h}q^2|\phi|^2(nA-\Phi)^2-n(2 p'-A'{\cal E})+\Phi{\cal E}'\biggr\rbrace+B
\end{eqnarray}
We will not write down the equations of motion but want to mention that,
even not completely obvious but nevertheless true, the reduced action vanishes
on-shell.

The action is invariant under
(\ref{dr}-\ref{dphi0},\ref{dp},\ref{dn}) and
\begin{eqnarray}
\tilde {\cal E}(\tilde r) & = & \sigma {\cal E}(r)  \label{dE}\,, \\
\tilde A(\tilde r) & = & \sigma A\label{dA}\,, \\
\tilde \Phi(\tilde r) & = & \sigma^{-1} \Phi(r)\label{dPhi}\,,\\
\tilde A_r(\tilde r) & = & \sigma^{-1} A_r(\sigma)\label{dAr}\,.
\end{eqnarray}
The conserved Noether charge is
\begin{eqnarray}
C & = & \gamma\Bigl[ h'+32 r h |\phi'|^2+16 iqrhA_r(\phi{\phi^*}'-\phi^*\phi')
+{16 h\over r}A'^2\Bigr]r\\
&&\qquad -n(2p'-{\cal E}A')r+\Phi{\cal E}' r-2\gamma h-{16\over r}h\gamma A A'
+4 n p-\Phi{\cal E}-n{\cal E}A\nonumber
\end{eqnarray}
Note that we have not yet fixed the gauge. We could set $A_r=0$ and $\Phi(r_+)=0$,
however for the following discussion only the latter gauge condition,
which ensures that the time component of the gauge field is regular at the horizon,
will be relevant.
We will, as before, also use $n(r_+)=0=h(r_+)$.

To derive the Smarr relation, we proceed as in the previous sections. Using the
scaling relations one derives
\begin{equation}\label{dBQ}
\delta B={1\over 8G}C\delta\sigma
\end{equation}
On the other hand
\begin{eqnarray}\
\delta B|_\infty&=&\beta(\delta M+\Omega\delta J+\Phi\delta Q)\\
&=&2\beta(M+\Omega J+{1\over2}\Phi Q)\delta\sigma\nonumber
\end{eqnarray}
from which we obtain
\begin{equation}\label{CQ}
C=16G\beta(M+\Omega J+{1\over 2}\Phi Q).
\end{equation}
The relative factor ${1\over2}$ comes from the fact that
$Q={\cal E}|_{r=\infty}$
has scaling charge one while $M$ and $J$ have scaling charge two.
We also have, as before
\begin{equation}
\delta B|_{r_+}=-\gamma\delta h|_{r_+}
\end{equation}
leading to
\begin{equation}
S={2\pi r_+\over 4G}
\end{equation}
Using that $C(r_+)=r_+\gamma_+ h'_+=4\pi r_+$ we finally find the
Smarr relation
\begin{equation}
{1\over2}S=\beta(M+\Omega J+{1\over2}\Phi Q)\,.\label{SmarrJQ}
\end{equation}

%===================================================
%===================================================
\section{D=4}
\label{TorusSec}

In four dimensions the equations of motion have a similar structure although there are
important differences. For reasons which will become clear very soon, we make the
general ansatz for the metric
\begin{equation}\label{4d}
ds^2 = -\gamma^2 h dt^2 + {dr^2 \over h} + r^2 d\Omega_k
\end{equation}
where the ``sphere"  $d\Omega_k$ is either a 2-sphere, a 2-torus or a higher genus
surface,
\begin{equation}
d\Omega_k = \left\{ \begin{array}{ll}
  d\theta^2 + \sin^2\theta d\phi^2, & k=1 \\
  dx^2 + dy^2, & k=0 \\
   d\theta^2 + \sinh^2\theta d\phi^2, & k=-1. \\
\end{array} \right.
\end{equation}
Black holes with unusual topologies were first discussed in \cite{Ami,Vanzo}.

The ansatz (\ref{4d}) leads to the reduced action
\begin{equation}\label{4daction}
I[h,\gamma,\phi] = -{(t_2-t_1)\over 8\tilde G}\int dr\, \gamma \left( rh' + h-k + 8r^2 h
\phi'^2 + 16 r^2 V(\phi) \right) +B\,.
\end{equation}
We have introduced the notation $\tilde G={4\pi G\over V_k}$ with $V_k=\int d\Omega_k$.
The horizon area is then $V_k r_+^2$. Varying $\gamma,h,\phi$ one obtains the equations
of motion
\begin{eqnarray}
  rh' + h - k + 8 r^2 h \phi'^2 + 16 r^2 V(\phi) &=& 0\,, \label{eq1} \\
  -\gamma' + 8 r \gamma \phi'^2 &=& 0\,, \label{eq2} \\
  -(r^2\gamma h \phi')' + r^2 \gamma V_{,\phi} &=& 0\,. \label{eq3}
\end{eqnarray}
These equations are similar to those in three dimensions, (\ref{3deq}), except for the
constant $k$ appearing in (\ref{eq1}). This constant, which is a fixed number associated
to the sphere's curvature, spoils scale invariance\footnote{Note that if one replaces
$\gamma=\lambda'$ and varies $\lambda$, the piece $\lambda' k$ is a boundary term and
the action becomes scale invariant. The space of solutions has an extra integration
constant, and for particular values of that constant, the original equations are
recovered. The main obstruction to follow up this idea is the relativistic version of
the modified equations of motion.}.

However, for the torus topology, $k=0$, the equations are scale invariant and we can
immediately generalize the discussion from $d=3$ to $d=4$\footnote{The generalization to
arbitrary $d$ is straightforward if we take for $d\Omega_{\rm d-2}$ the volume element
of a flat torus.}, in particular due to scale invariance there is a radially conserved
charge. Due to the invariance of the action under the replacements
$(r,h(r),\gamma(r),\phi(r))\to (\tilde r,\tilde h(\tilde r), \tilde\gamma(\tilde
r),\tilde\phi(\tilde r))$ with ({\it c.f.} (\ref{dr}-\ref{dphi0}))
\begin{eqnarray}
   \tilde r &=& \sigma r \label{dr40}\\
  \tilde h(\tilde r) &=& \sigma^2 \, h(r)  \label{dh40}\\
  \tilde \gamma(\tilde r) &=& \sigma^{-3} \gamma(r) \label{dg40}\\
  \tilde \phi(\tilde r) &=& \phi(r)  \label{dphi40} \ \ \ \
\end{eqnarray}
one finds
\begin{equation}\label{C40}
C=\gamma\left(r^2 h'-2h+8 r^3 h\phi'^2\right)
\end{equation}
with $C'=0$ by virtue of the equations of motion.

Eq.(\ref{S}) for the entropy is now
\begin{equation}\label{Sd=4}
S={4\pi r_+^2\over 4 \tilde G}={V_0 r_+^2\over 4 G}\,.
\end{equation}
and
\begin{equation}\label{dM40}
\delta M=-\left.{V_0\over 8\pi G}\left(r\delta h+8 r^2 h
\phi'\delta\phi\right)\right|_{r=\infty}
\end{equation}
Using (\ref{dphi40},\ref{dh40}) and the fact that $M$ now has scaling weight three, one
finds from (\ref{dM40}) and from comparing with (\ref{C40}) the relation
\begin{equation}\label{CM4}
C={24\pi G\beta\over V_0}M\,.
\end{equation}
On the other hand, evaluation of $C$ at the horizon gives
\begin{equation}\label{C4hor}
C=4\pi r_+^2\,.
\end{equation}
Comparison of (\ref{CM4}) with (\ref{C4hor}) leads to the relation
\begin{equation}\label{rel4}
\beta M={V_0 r_+^2\over 6 G}\,.
\end{equation}
In place of (\ref{kappa}) one now finds
\begin{equation}\label{kappa4}
8\pi GM=V_0 \kappa_0^3 r_+^3
\end{equation}
and the specific heat can be computed to be twice the entropy and for the temperature
one finds
\begin{equation}\label{T40}
T={3\over2}\pi^{-2/3}\kappa_0^2\left({GM\over V_0}\right)^{1/3}\,.
\end{equation}
We stress once more that these  results are valid for arbitrary potentials as long as
they lead to a solution for the scalar field which vanishes asymptotically. The specific
form of the potential only enters through the integration constant $\kappa_0$.

The proof of positivity of $M$ proceeds in exactly the same way as in $d=3$. It depends
crucially on the existence of the scaling charge, i.e. on considering the case $k=0$. In
fact, negative mass hairy black holes for $k=1$ have recently been constructed in
\cite{HeHo2}.

It is now straightforward to check that for a constant potential $V=-3$, i.e. in the
presence of a cosmological constant but no scalar field, one finds the above results
with $\kappa_0=1$ as one easily verifies given the explicit solution
 \begin{equation}
 h = r^2-{8\pi Gm \over V_0 r}
 \end{equation}

\newpage

\section{Acknowledgments}

The authors gratefully acknowledge comments on the manuscript by S. Carlip and G.
Horowitz, and G. Clement for pointing out an important misprint in the first version. We
would also like to thank the Centro de Estudios Cient\'\i ficos in Valdivia, Chile and
in particular J. Zanelli for hospitality during the initial stages of this work. M.B.
was partially supported by FONDECYT (Chile) grants \# 1020832 and \# 7020832. M.B. also
thanks the Albert Einstein Institute, Potsdam,  and ULB, Brussels for their kind
hospitality during the course of this work, and G. Barnich and P. Liendo for useful
conversations. S.T. thanks the Physics Departments of P. Universidad Cat\'olica de Chile
and of the Universidad Aut\'onoma in Madrid for hospitality, and A. Schwimmer and M.
Volkov for useful discussions. His work is partially supported by GIF -- the
German-Israeli Foundation for Scientific Reasearch and the EU-RTN network {\it
Constituents, Fundamental Forces and Symmetries of the Universe} (MRTN-CT-2004-005104).

 \end{document}